\documentclass{elsart}
\usepackage{psfig,natbib}
\begin{document}

\def\approxlt{\mathrel{\hbox{\rlap{\lower.55ex \hbox {$\sim$}}
        \kern-.3em \raise.4ex \hbox{$<$}}}}
\def\approxgt{\mathrel{\hbox{\rlap{\lower.55ex \hbox {$\sim$}}
        \kern-.3em \raise.4ex \hbox{$>$}}}}
\def\ros{{\sl ROSAT }}
\def\asca{{\sl ASCA }}
\def\G{$\Gamma_{\rm x}$ }

\runauthor{St. Komossa et al.}
\begin{frontmatter}
\title{1E 0117.2-2837 and the NLR of Narrow-Line Seyfert\,1 galaxies}
\author[s]{Stefanie Komossa},
\author[s]{Dirk Grupe},
\author[j]{ M. Janek}
\address[s]{Max-Planck-Institut f\"ur extraterrestrische Physik, Giessenbachstr., D-85748 Garching, Germany;
email: skomossa@xray.mpe.mpg.de}
\address[j]{D-31683 Obernkirchen} 
\begin{abstract}
Based on photoionization calculations we present a study 
of the NLR emission line ratios of Narrow-line Seyfert 1 (NLS1) galaxies.
In particular, we investigate the influence of different EUV - soft-X-ray spectral shapes
(a giant soft excess, a steep X-ray powerlaw, the presence of a warm absorber)
and NLR cloud properties (density, abundances, distance from the continuum source)
on the predicted optical emission-line ratios like [OIII] and [FeXIV]. 
\end{abstract}
\end{frontmatter}

\section{Introduction}
Narrow Line Seyfert 1 (NLS1) galaxies have recently received 
much attention
due to their unusual optical--X-ray properties
which are not yet well understood.
Photoionization models of the circum-nuclear emission/absorption regions
allow us to investigate
scenarios to explain the
main characteristics of NLS1s, i.e., (i) extremely steep X-ray spectra within the \ros energy band,
(ii) narrow (FWHM $<$ 2000 km/s) Balmer lines and strong FeII emission, 
and (iii) weak forbidden lines except for some relatively strong high-ionization iron lines.
Here, we concentrate on (iii); for a discussion of (i) and (ii)
see Komossa \& Fink (1997a), and Komossa \& Meerschweinchen (2000).
In particular, we study the influence of different EUV - soft-X-ray spectral shapes
(a giant soft excess, a steep X-ray powerlaw, the presence of a warm absorber)
and NLR cloud properties (density, abundances, and distance from the nucleus)
on the predicted optical emission-line ratios.
First results of this study were presented by Komossa \& Janek (1999).

\section{1E 0117.2-2837} 
1E 0117.2-2837 (QSO\,0117-2837) was discovered as an X-ray source by
{\sl Einstein} and is
at a redshift of $z$=0.347 (Stocke et al. 1991, Grupe et al. 1999).
Its X-ray spectrum is extremely steep (see Komossa \& Meerschweinchen 2000
for a detailed X-ray analysis of this source). 
We have obtained new optical spectra of 1E 0117.2-2837 
with the ESO\,1.52\,m telescope at LaSilla, in September 1995. 
The optical spectrum reveals several signs of a
NLS1 galaxy (we do not distinguish between NL Seyferts
and NL quasars, here):
weak [OIII]$\lambda$5007 emission and strong
FeII complexes (Fig. 1).
After subtraction of the FeII spectrum (see Grupe et al. 1999 for details)
we derive FWHM${_{\rm H\beta}}$=2100$\pm{100}$ km/s,
FWHM${_{\rm [OIII]}}$=820$\pm{150}$ km/s
(based on {\em single-component} Gaussian fits to the emission
lines; this leaves some broad wings as residuals),
and the ratio [OIII]/H$\beta$=0.056.

\begin{figure}[hb]
\hspace*{2cm}
{\psfig{file=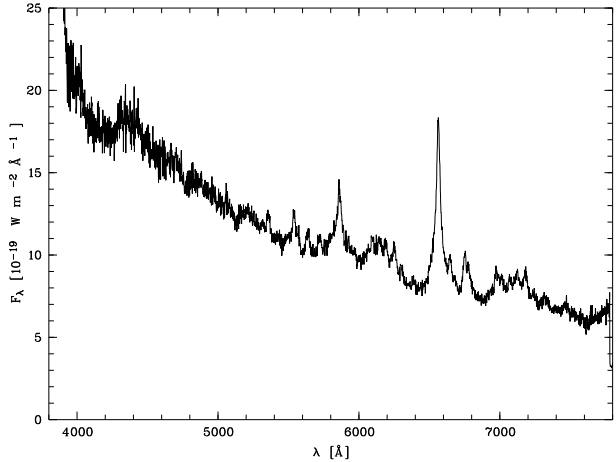,width=9.4cm,angle=-90,clip=}}
\caption[]{Optical spectrum of QSO\,0117-2837 taken with the ESO 1.5m telescope
at LaSilla. The strongest line is that of H$\beta$.}
\end{figure}

\section{The NLR of NLS1 galaxies: results from photoionization modeling}
The narrow emission lines, i.e. those originating from the narrow line region (NLR),
like [OIII]$\lambda$5007 and [OI]$\lambda$6300,
are rather weak in 1E 0117.2-2837 and in NLS1 galaxies in general.
Occasionally, however, fairly strong high-ionization iron lines are present.
We investigate several models to explain these
observations, starting with the assumption that the NLR is `normal'
(i.e. a typical type 1 Seyfert, as far as distance
from the nucleus, gas density and covering factor are concerned).
All photoionization calculations were carried out using 
the {\em{Cloudy}} code (Ferland, 1993). 

In the first step, non-solar metal abundances were assumed.
Over abundant metals (with respect to
the solar value) were shown to delay the complete removal of a BLR multiphase
equilibrium (Komossa \& Meerschweinchen 2000). Due to their rather strong influence on the cooling,
metals, if overabundant,
can lead to weaker optical line emission.
However, we find the effect to be insufficient to explain
the observed line intensities.

As shown in Komossa \& Schulz (1997), the weak [OIII]$\lambda$5007 domain
of the line correlations in the usual diagnostic diagrams of Seyfert 2
galaxies can be explained by very steep EUV continua with $\alpha_{\rm
uv-x}$ $\approx$ --2.5. Although, e.g., the NLS1 RX\,J1225.7+2055 indeed
exhibits a rather {\em steep} EUV spectrum (determined by a powerlaw
connection between the flux at the Lyman limit and 0.1 keV), that of
RX\,J1239.3+2431 is very {\em flat} (Greiner et al. 1996).

Placing warm absorbing material along the line of sight to the NLR would make
the latter see a continuum that is only modified in the soft X-ray region, with
negligible influence on the line emission. The same holds for an intrinsically
steep X-ray powerlaw, which only leads to a slight weakening of the low-ionization lines.

In cases where a warm absorber is present, the high-ionization iron lines
([FeX] and higher) can be produced within the warm gas itself (see
Komossa \& Fink 1997a,b,c for details). However, no one-to-one match
between the observed coronal lines in the NLS1 NGC\,4051 and those
predicted to arise from the X-ray warm absorber in this galaxy was found
(Komossa \& Fink 1997a, and these proceedings), suggesting that in
general, the coronal line and warm absorber regions are separate components.

In order to assess the influence of a strong EUV - soft-X-ray excess
on optical line emission,
we have calculated a sequence of models with an underlying mean Seyfert continuum
plus a black body of varying temperature,
for a range of densities and distances of the NLR gas from the central continuum source.
Although the contribution of a hot bump-component can considerably strengthen the
high-ionization iron lines (eg. [FeX]$\lambda$6374, [FeVII]$\lambda$6087
and  [FeXIV]$\lambda$5303), reflecting the fact that their
ionization potentials are at soft X-ray energies,
these models overpredict the [OIII] emission.

We conclude that the weakness of forbidden lines in NLS1s must be due to an overall
lower emissivity of the NLR.
If this is caused by {\em shielding} of the NLR from ionising photons,
the model must avoid boosting 
the low-ionization lines like [OII].
More likely, the region is gas poor, i.e. less of the impinging photons can be
reprocessed into line emission.
`Normal' line ratios would then result in weak forbidden lines being undetected.
For those objects with strong high-ionization (iron) lines,
models favor the dominance of low-density
gas and small distances to the ionizing source.

\begin{figure}[ht]
{\psfig{file=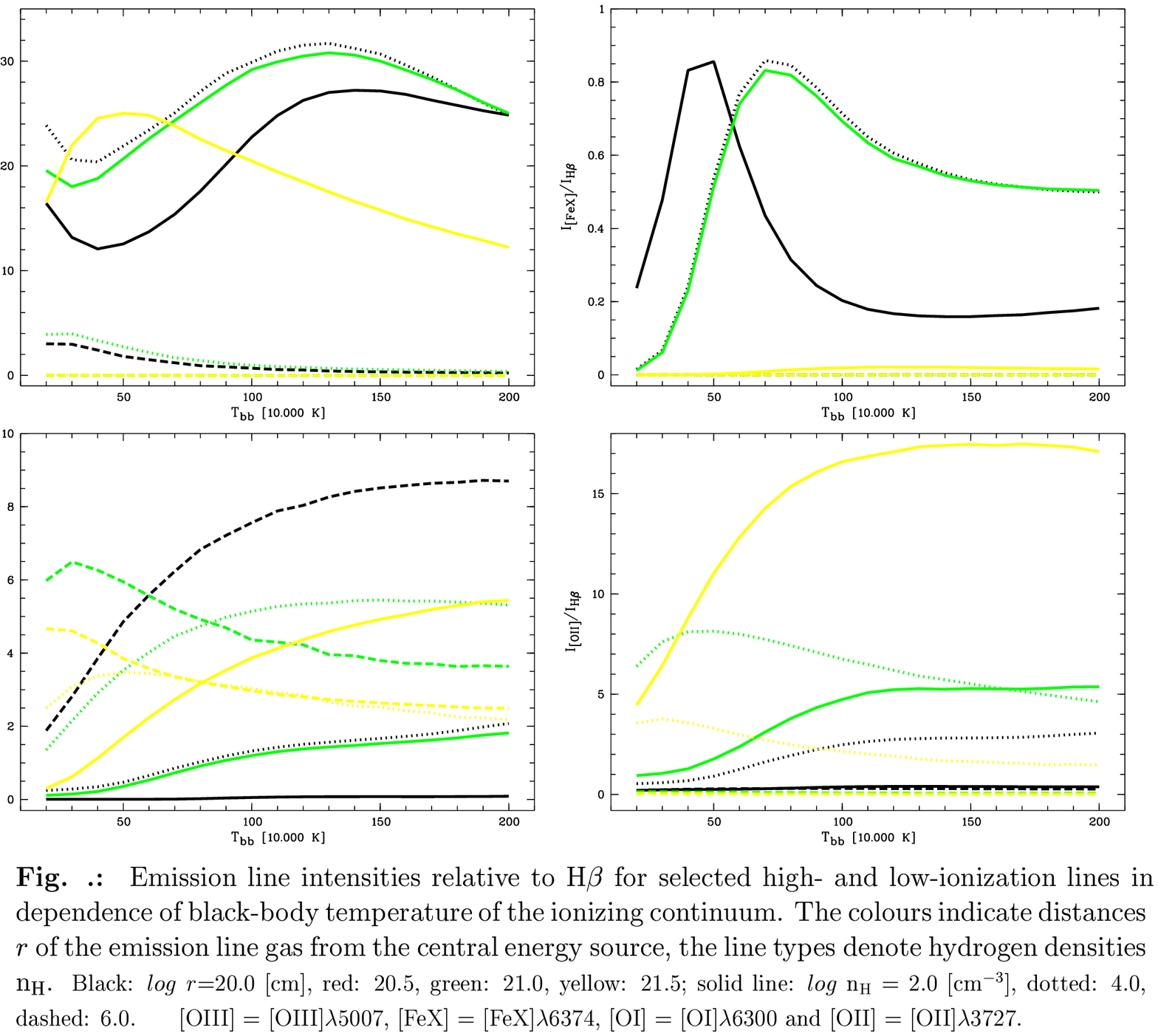,width=13.2cm,angle=90,clip=}}
\caption[]{Emission line intensities relative to H$\beta$ for selected
 high- and low-ionization lines in dependence of black-body temperature $T_{\rm bb}$ of the
ionizing continuum (see text for details). The grey-scales indicate distances $r$ of
 the emission line gas from the central energy source,
the line types denote hydrogen densities n$_{\rm H}$.
Black: $log$ $r$=20.0 [cm],
dark grey: 21.0, light grey: 21.5;  solid line: $log$ n$_{\rm H}$ = 2.0 [cm$^{-3}$], dotted: 4.0,
dashed: 6.0. [OIII] = [OIII]$\lambda$5007, [FeX] = [FeX]$\lambda$6374, [OI] = [OI]$\lambda$6300
and [OII] = [OII]$\lambda$3727.}
\end{figure}

%
\noindent {\sl Acknowledgements:}
\noindent We thank Gary Ferland for providing {\em Cloudy}.
Preprints of this and related papers can be retrieved from our webpage
at \\
http://www.xray.mpe.mpg.de/$\sim$skomossa/

\end{document}